\documentclass[12pt,a4paper,final]{article}
\usepackage[utf8]{inputenc}
\usepackage[english]{babel}
\usepackage[ final ]{pdfpages}
\usepackage{amsmath}
\usepackage{braket}
\usepackage{physics}
\usepackage{amsfonts}
\usepackage[english]{babel}
\usepackage{amssymb}
\usepackage{graphicx}
\graphicspath{ {images/} }
\author{\textbf{Renzo Mosetti}\footnote{rmosetti@ogs.trieste.it}\\\ University of Trieste\\\
\small \textit{Department of Mathematics and Geoscience, Via Weiss, 4, 34127, Trieste, ITALY}}
\title{Principal Component Analysis\\of Quantum Correlation}
\begin{document}
\maketitle
\textbf{Abstract}.The concept of quantum correlation matrix for observables leads to the application of the PCA (Principal Component Analysis) also for quantum system in Hilbert space. It is shown that, in the case of a 2x2 spin system where the observables are the Pauli matrices and a polarization vector is acting on the density matrix, meaningful results can be obtained by the PCA in terms of qubit properties. In particular, it is shown that by choosing, for example, the (x,z) spinors, the axes of the principal components (PC) coincide with the circular polarization basis on the Bloch Sphere.

\section{Introduction}

The interest in quantum computation has raised a huge number of papers  concerning the statistical properties of quantum systems. In particular the concept of quantum Principal Component Analysis (qPCA) has been defined \cite{lloyd} . However, in the context of this paper, the qPCA is introduced as an algorithm for quantum tomography. Quantum tomography is the process of obtaining features of an unknown quantum state defined by a density matrix $\varrho$. The authors make use of the standard PCA on multiple copies of an unknown density matrix to construct the eigenstates corresponding to the large eigenvalues of the state. In practice, the qPCA is a method to derive, in a efficient computational way , the eigenvalues and eigenvectorrs of $\varrho$. Actually, in analogy of what happens in the classical correlation analysis, no definition of PCA in terms of quantum correlation among observables has been derived. The aim of this paper is to show that it is possible, starting from a possible definition of a quantum correlation function, to derive a PCA following the classical development of the standard correlation analysis. 
 
\section{Quantum Correlation Matrix and PCA}
A proposed definition of a quantum mechanical correlation function has been developed in the framework of the classical limit of quantum mechanics \cite{ana}. By different considerations, the same correlation function has been suggested by Andersen in an interesting lecture note \cite{andersen}. The definition is based on the well known concept of average and variance of quantum observables within the Heisenberg picture. It is a reasonable definition for quantum mechanical correlation and has many of the properties of classical equilibrium correlation functions. In this paper the definition by Andersen is accepted and, as will be more clear later, this choice give rise to interesting developments in terms of application also to quantum computing. In this paper we are considering for our purposes only the correlation at zero time lag yielding the usual correlation matrix. 

Let $O_{1}; O_{2};....;O_{n}$ a set of observables. Giving the density matrix $\varrho$ describing the quantum system under consideration, the variance of an observable (assumed or reduced to a zero average value) $O_{i}$ is given by:
\begin{equation}
var (O_{i}) = tr(\rho O_{i}^2)
\end{equation}
In a similar way, the quantum correlation $ S_{O_{i}O_{j}}$ between two observable $O_{i}$ and $O_{j}$ is defined as:
\begin{equation}
 S_{O_{i}O_{j}} \equiv tr(\rho O_{i}O_{j})
\end{equation}
Now, it is possible to construct the Correlation Matrix $S$ in the same way as for the classical correlation analysis:
\begin{equation}
\\\\S\equiv
\begin{pmatrix}
tr(\rho O_{1}^2) &  tr(\rho O_{1}O_{2}) & .... & tr(\rho O_{1}O_{n})\\ 
tr(\rho O_{2}O_{1}) & tr(\rho O_{2}^2)& ....& tr(\rho O_{2}O_{n})\\
....&....&....&....\\
tr(\rho O_{n}O_{1})&....& .....&tr(\rho O_{n}^2)\\
\end{pmatrix}\\\\\
\end{equation}

\textbf{Remark}. The matrix $S$ could be non-Hermitian since, contrary of what happens in the ordinary correlation, the elements $tr(\rho O_{i}O_{j}), 
tr(\rho O_{j}O_{i})$ could be not the same due to the fact that operators do not in general commute. However, it is possible to define: $ \phi_{AB}=1/2(S_{AB} + S_{BA})
$
then the function $\phi$ has the same properties of the classical correlation function.

At this point a question could be raised. Is it possible to extend in a formal way the Principal Components Analysis (PCA) to the quantum correlation matrix so defined? If the answer is yes, what is the meaning of the Principal Components for a quantum system. As a first insight, the first PC constitutes a new coordinate axis in the variable space which is oriented in order to maximize the variations of the projections of the data points on this axis. It is clear that, within the quantum mechanics context, the projection is the process of measuring an observable and then, the first PC may represent the measurement of an observable along the axis of the maximum of the variance of the measurements. So, this seems to be reasonable and potentially useful also for quantum systems.
\subsection{The PCA theorem}
We recall that the PCA \cite{book} are defined within an Euclidean space $ R^{n}$ as the problem to find the maximum of the function $var(z)=a^{T}S a \equiv V(a)$ where:
$a\in R^{n}; z=a^{T}x; $ and being S the covariance (correlation) matrix of the n dimensional vectors of variables x. 
In the case, where observables are matrices (let's consider only the finite dimensional case), one has to move to the space:
$ M(n,C) $ being n the order of the matrix. 
The problem becomes : find the maximal variance, as defined in (1), of a matrix P expressed as a linear combination of observable matrices where the elements of the combination are complex parameters $a_{i} $ subjected to the constraint: $ \Sigma_{i}a_{i}^{*}a_{i}=1$
 
\subsubsection{ THEOREM:  Principal Components on a Matrix Space}
The maximum of the  variance of a Matrix $P$ expressed as:

$P=a_{1}O_{1} + a_{2}O_{2}+....+a_{n}O_{n}$
with the constraint 
$ \Sigma_{i}a_{i}^{*}a_{i}=1 $
is equal to the maximum of the eigenvalues $\lambda_{1}$ of the covariance matrix $S$ and the n coefficients $a_{i}$ are the components of the complex vector which is the eigenvector $a_{1}$ of $\lambda_{1}$.
 
PROOF.
Let's define the function $F(\textbf{a}) = tr[(\varrho(a_{1}O_{1}+a_{2}O_{2}+..+a_{n}O_{n})^{2}]+\lambda(1-a^2_{1}-a^2_{1}-...-a^2_{n})$
where$ \lambda$ are the Lagrange multipliers related to the constraint for the maximization problem.   The necessary conditions for the maximum are;
\bigskip

$\partial F/\partial a_{1}=0; \partial F/\partial a_{2}=0; ..\partial F/\partial a_{n}=0;$
\bigskip

By using the properties of the trace, we obtain:
\bigskip

$\partial F/\partial a_{i}= 2a_{i} tr(\rho O_{i}^2) + 2\Sigma_{j=1,i\neq j}a_{j}tr(\rho O_{i}O_{j}) -2\lambda a_{i} = 0$
\bigskip

By rearranging the above expression we get:
\bigskip

$a_{i} tr(\rho O_{i}^2) + \Sigma_{j=1,i\neq j}a_{j}tr(\rho O_{i}O_{j}) = \lambda a_{i}$
\bigskip

For $i=1,n$ the following eigenvalue problem is obtained in compact form:
\medskip

$S \textbf{a}=\lambda\textbf{a}$
\bigskip

where:
\bigskip

$S=
\begin{pmatrix}
tr(\rho O_{1}^2) &  tr(\rho O_{1}O_{2}) & .... & tr(\rho O_{1}O_{n})\\ 
tr(\rho O_{2}O_{1}) & tr(\rho O_{2}^2)& ....& tr(\rho O_{2}O_{n})\\
....&....&....&....\\
tr(\rho O_{n}O_{1})&....& .....&tr(\rho O_{n}^2)\\
\end{pmatrix}$
\bigskip

Note that $S$ is exactly the quantum covariance matrix as defined in (3).
This gives the proof of the theorem.
\section{Application to a 2x2 spin system}
	In order to show in a specific case what could be the potentialities of the extension of the PCA to a set of observables by using the quantum correlation matrix, an application will be given for a 2x2 spin system. This leads to interesting implications to quantum computing, being the 2x2 spin system the usual qubit representation space. Let's start  from the observables that in this case are the Pauli matrices:
\bigskip

$
\sigma_{x}=
\begin{pmatrix}
0 & 1 \\ 
1 & 0
\end{pmatrix};
\sigma_{y}=
\begin{pmatrix}
0 & -i \\ 
i & 0
\end{pmatrix};
\sigma_{z}=
\begin{pmatrix}
1 & 0 \\ 
0 & -1
\end{pmatrix}$
\bigskip

The general form of the density matrix $\rho$ where a polarization vector$\\ \textbf{P}=(Px,Py,Pz)$ is acting on the system, is:
\bigskip

$
\rho=1/2
\begin{pmatrix}
1+Pz & Px-iPy \\ 
Px+iPy & 1-Pz
\end{pmatrix}$
\bigskip

For pure state$\ |\textbf{P}|=1 $ while for mixed states $|\textbf{P}|< 1.$
Now, for any pairs of Pauli matrices, the elements of the correlation matrix are:
\medskip
\begin{equation}
S_{i,j}= tr(\rho \sigma_{i}\sigma_{j})\hspace{0.2cm} for \hspace{0.3cm}  i=x,y,z;j=x,y,z; i\neq j
\end{equation}
\pagebreak

Consider, for instance :
\medskip

$
S_{x,z}=
\begin{pmatrix}
1 & -iPy \\ 
iPy &  1
\end{pmatrix}$
\medskip

The PCA are obtained by solving the eigenvalue problem:
\medskip

$
S\ket{\phi} = \lambda \ket{\phi}. $
\medskip

Explicit computation for $S_{x,z}$ gives:
\medskip

$ \lambda_{1}=1+Py;\hspace{1cm} \lambda_{2}= 1-Py $
\medskip

The corresponding eigenvectors are (after a normalization) :
\begin{equation}
\ket{\phi_{1}} = 
\begin{matrix}%
(-i\sqrt{2}, &1/\sqrt{2})^{T}
\end{matrix};\\\\
\ket{\phi_{2}} = 
\begin{matrix}
(i\sqrt{2}, &1/\sqrt{2})^{T}
\end{matrix}\\
\end{equation}
This process is equivalent to find the coordinate system in which the correlation matrix is diagonal. The eigenvector with the largest eigenvalue ($\lambda_{1}$) is along the direction of the highest variance, the second  eigenvalue ($\lambda_{2}$) corresponds to the lowest variance and the relative eigenvector lies in an orthogonal direction. 
Now, let us express the eigenvectors in the so called standard computational basis on the Bloch Sphere:
\medskip

$
\ket{0} = 
\begin{matrix}
(1,&0)^{T}
\end{matrix};\ \hspace{1cm}
\ket{1} = 
\begin{matrix}
(0,&1)^{T}
\end{matrix}\\
$
\medskip

The new basis vectors that can be derived from (5) are:
\medskip

$\ket{0'} = 
1/\sqrt{2}(\ket{0}+i\ket{1}) \hspace{1cm}    \ket{1'} = 1/\sqrt{2}(\ket{0}-i\ket{1})$
\medskip

The principal axes obtained by PCA analysis in this case correspond  to the choice of the circular polarization basis for qubits which is a well known basis and it is used for quantum gates.
The PCA are :
\bigskip

$P_{1}=-i/\sqrt{2}\begin{pmatrix}
0 & -i \\ 
i & 0
\end{pmatrix} + 1/\sqrt{2}\begin{pmatrix}
0 & -i \\ 
i & 0
\end{pmatrix}\\\
P_{2}=i/\sqrt{2}\begin{pmatrix}
0 & -i \\ 
i & 0
\end{pmatrix} + 1/\sqrt{2}\begin{pmatrix}
0 & -i \\ 
i & 0
\end{pmatrix}$
\bigskip

It is easy to show that $P_{1}$ and $P_{2}$ are uncorrelated according to definition (2). The meaning of the PCA matrices follows from the computation of the variance as defined in (1). Explicit computation gives:
\bigskip

$
var(P_{1}) = 1 + P_{y}; \hspace{1cm}   var(P_{2}) = 1 - P_{y}
$
\bigskip

This is fully consistent with the PCA since the variances correspond to the eigenvalues  $\lambda_{1}$ and $\lambda_{2}$ obtained by the eigenvalue problem. The maximum of the variance (by assuming $P_{y}>0$) is contained in the first Principal Component $P_{1}$ and the value is that of the corresponding eigenvalue $\lambda_{1}$.
\section{Conclusions}
The idea behind this paper was to investigate the possibility to prove that a definition of quantum correlation between observables leads to further improvements in quantum statistical properties. In particular it has been shown that the Principal Component Analysis can be performed on the quantum correlation matrix. The obtained results seem to be very fruitful and open the door to some applications to quantum computing. The fact to project quantum states on axes of maximal variance may be used for instance in quantum data compression. In a very interesting paper on this subject \cite{schure} the application of the Schur-Weyl transform is used to compress an ensemble of identically prepared qubits.It is well known, from a mathematical point of view, that relationships exist between Singular Value Decomposition, PCA \cite{pca} and the Schur-Weyl theory \cite{math}. These aspects have to be investigated in more detail in the framework of quantum statistics and quantum computing.

\end{document}